\documentclass[twocolumn,showpacs,prl,superscriptaddress]{revtex4}
\usepackage{graphicx}
\usepackage{color}\usepackage{enumerate}
\usepackage{amsmath}\usepackage{amssymb}\usepackage{stmaryrd}
\usepackage{esint}
\usepackage{multirow}
\usepackage{float}
\usepackage{longtable,booktabs}
\usepackage{subfigure}
\usepackage{dsfont}
\usepackage{amssymb}
\usepackage{comment}
\usepackage{bm}
\usepackage[pdftex,colorlinks=true,linkcolor=blue,citecolor=blue]{hyperref}

\usepackage{mathtools}

\DeclarePairedDelimiterX\braket[2]{\langle}{\rangle}{#1 \delimsize\vert #2}

\begin{document}
\title{Higher Rank Chiral Fermions in 3D Weyl Semimetals}
\author{Oleg Dubinkin}
\affiliation{Department of Physics and Institute for Condensed Matter Theory, University of Illinois at Urbana-Champaign, Illinois 61801, USA}

\author{F.~J. Burnell}
\affiliation{Department of Physics, University of Minnesota Twin Cities, 
MN, 55455, USA}

\author{Taylor L. Hughes}
\affiliation{Department of Physics and Institute for Condensed Matter Theory, University of Illinois at Urbana-Champaign, Illinois 61801, USA}
\begin{abstract}
We report on exotic response properties in 3D time-reversal invariant Weyl semimetals with mirror symmetry.  Despite having a vanishing anomalous Hall coefficient, we find that the momentum-space quadrupole moment formed by four Weyl nodes determines the coefficient of a mixed \emph{electromagnetic charge-stress} response, in which momentum flows perpendicular to an applied electric field, and electric charge accumulates on certain types of lattice defects.  This response is described by a mixed Chern-Simons-like term in 3 spatial dimensions, which couples a rank-2 gauge field to the usual electromagnetic gauge field.   On certain 2D surfaces of the bulk 3D Weyl semimetal, we find what we will call rank-2 chiral fermions, with $\omega =k_x k_y$ dispersion. The intrinsically 2D rank-2 chiral fermions have a mixed charge-momentum anomaly which is cancelled by the bulk of the 3D system. \end{abstract}

\maketitle

Chiral fermions have had a remarkable impact across a variety of fields of physics in the past few decades. Whether it is in the context of the weak interactions in particle-physics\cite{LeeYang,MadameWu}, or as low-energy edge or bulk excitations of topological insulators\cite{halperin,haldane1988} and semimetals\cite{nielsen1983,Wan2011,armitage2018}, or even as heralds of non-reciprocal light transport in photonic crystal analogs\cite{Haldane08,Raghu08,ozawa2019}, there is no denying their broad relevance to a number of physical platforms. From their origin in Lorentz-invariant field theories it is known that these massless, linearly dispersing fermions can intrinsically appear in any \emph{odd} spatial dimension. Additionally, in a condensed matter context, the famous Nielsen-Ninomiya no-go theorem\cite{Nielsen81} dictates that local, time-independent lattice Hamiltonians must harbor an even number of chiral fermions, such that the total chirality vanishes. Hence, these restrictions allow 1D chiral fermions to either appear as right/left-mover pairs in a 1D metal, or as isolated chiral edge states of a Chern insulator\cite{haldane1988}, while 3D chiral (Weyl) fermions appear in nodal pairs in Weyl semimetal materials\cite{nielsen1983}.

Recent developments in the condensed matter and high-energy literature have opened the door to the discovery of new types of massless fermions and bosons in a non-Lorentz invariant, crystalline environment\cite{paramekanti2002,cano,hourglass,seiberg1,seiberg2,seiberg3}. In this work we propose a generalization of chiral fermions for 2 dimensional systems with crystalline symmetry; we call these rank-2 chiral fermions.  
We present a 2D lattice model exhibiting rank-2 chiral fermions, but where the total higher rank chirality vanishes. Then we present a 3D model in which the rank-2 chiral fermions appear as surface states with an associated anomalous response.  We find the anomalous response can be mapped onto a bulk rank-2 Chern-Simons term (in analogy with the dipole Chern-Simons term from Ref. \onlinecite{ybh2019}) representing a mixed charge-geometric response, and has a coefficient related to the Berry curvature quadrupole moment of the bulk Fermi-surface.

We begin by reviewing chiral fermions in 1D.  For gapless fermions in 1D, the chirality is given by $\chi_1={\rm{sgn}}(v)$, where $v$ is a characteristic velocity that fixes the dispersion relation $E(k)=\hbar vk- \mu$.  Superficially, this chirality allows us to define {\it two} currents that are conserved by the classical equations of motion of a general 1D system: the usual charge current $j^{\mu}$
(where $j^0 = \sum c^\dag_{\alpha}  c_\alpha$), and the axial current $j^{\mu}_\chi$ (where $j^0_\chi = \sum_\alpha \chi_{\alpha}  c^\dag_{\alpha}  c_{\alpha} $), and $\alpha$ runs over all fermion channels.  Evidently, $j^0_\chi$ is associated with the conservation of the difference in the number of right-movers (with positive chirality) and the number of left-movers (with negative chirality).

However, it is well known that these two currents cannot be simultaneously conserved\cite{Adler,BellJackiw,Peskin}. Indeed, in the presence of an electric field $E_x$, the axial current obeys the anomalous conservation law
\begin{equation} \label{Eq:An1d}
    \partial_\mu j^\mu_{\chi}=n \frac{e E_x}{\pi \hbar},
\end{equation}\noindent where $n$ is the number of channels.  
This anomalous response reflects the fact that during a process in which one adiabatically shifts the vector potential $A_x\to A_x+\frac{h}{eL},$ the number of right-moving (left-moving) particles changes by $\delta N_{\chi_1=+1} = +1$ ($\delta N_{\chi_1=-1} = -1$).  Strikingly, if $\sum_\alpha \chi_\alpha \neq 0$, the axial anomaly (\ref{Eq:An1d}) also implies an anomaly in the U(1) charge current.   Thus a net chirality is impossible in a 1D lattice system that conserves the electric charge.  However, chiral fermions can appear on the 1D edges of 2D integer quantum Hall \cite{halperin,wen1992} or quantum Anomalous Hall \cite{haldane1988,2013paper} systems. Moreover, a whole family of quasi-1D chiral edge states exists on the boundary of 3D, T-breaking Weyl semimetals, and forms a so-called Fermi arc having chiral dispersion in the surface Brillouin zone\cite{Wan2011,armitage2018}.  In these cases, the anomalous conservation law on one boundary is balanced by a current flow through the bulk (possibly from the other boundary)\cite{callanharvey}, and is a signature of a Hall effect.

\begin{figure}
    \centering
    \includegraphics[width=0.5\textwidth]{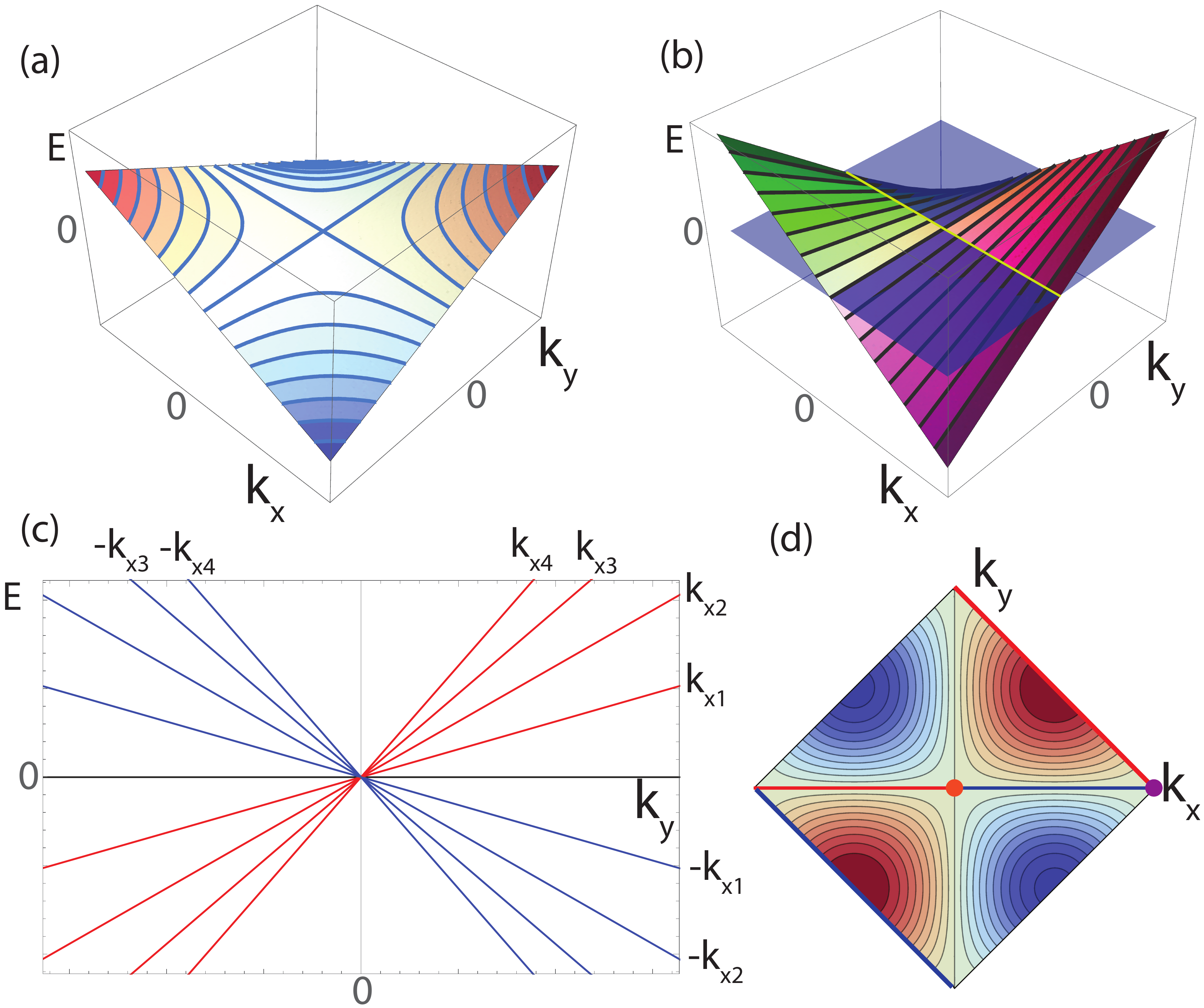}
    \caption{(a) Dispersion relation and Fermi surface contours for Eq. \ref{kxky}. Note that when $\mu=0$, the the Fermi surface consists of two intersecting lines. (b) Dispersion relation of Eq. \ref{kxky} but with guides to illustrate the chirality $\chi^{\hat{y}}_{K_x}$ in the $k_y$-direction for fixed values of $k_x.$ The yellow line represents the collection of $k_{\parallel}$ Fermi points at a fixed Fermi level represented by the blue plane. (c) Cross-sections of the dispersion as a function of $k_y$ for a set of fixed values of $k_x$, colors indicate different $\chi^{\hat{y}}_{K_x}$ chiralities which are opposite for $\pm k_x$. (d) Fermi-level contours for the 2D tightbinding dispersion $E({\bf{k}})=\sin k_x \sin k_y.$  Orange and blue points are rank-2 chiral fermions having $\chi_2=\pm 1$ respectively. Thick red/blue lines indicate regions of the $\mu=0$ Fermi surface that gain/lose particles when an electric field $E_x$ is turned on. Globally there is no momentum anomaly as each fixed $k_x$ slice has a positive and negative chirality. The same is true for $E_x\to E_y, k_x\to k_y$ if we rotate the thick red and blue lines by $\pi/2$ counterclockwise.}
    \label{fig:3dplots_rnak-12}
\end{figure}

We now discuss a generalization of chiral fermions to 2D.  Consider a portion of a 2D Fermi surface, described locally by the dispersion relation:
\begin{equation} \label{kxky}
    E(\textbf{k})= \hbar v\xi k_x k_y - \mu,
\end{equation} where $v$ is a velocity and $\xi$ has units of length. Such local Fermi-surface patches are not uncommon and indeed are guaranteed in 2D bands by Morse theory\cite{van1953}.
The dispersion relation and Fermi surface contours are depicted in Fig. \ref{fig:3dplots_rnak-12}(a).  To look for a charge anomaly we can consider the current $j^i=\partial H/\partial k_i = \hbar v\xi \sigma^{ij}k_j$ where $\sigma^{xy}=\sigma^{yx}=1,\sigma^{xx}=\sigma^{yy}=0.$ We see that upon adiabatically turning on a constant electric field $E_x$ or $E_y$, the total number of charged particles below the Fermi surface does not change: for every extra fermion that is added at momentum $\sigma^{ij}k_j$ in the presence of the electric field $E_i$, there is a partner at the opposite momentum $-\sigma^{ij}k_j$ that is removed.  This should be contrasted with the response for a 1D chiral fermion above where inserting a single flux of $E_x$ generates an extra particle, leading to an anomaly in the U(1) charge current; such an anomaly \emph{is not} present in our rank-2 chiral system. However, we will now show that the dispersion relation (\ref{kxky}) does exhibit a 2D variant of the axial anomaly, which leads to a violation of momentum conservation in the presence of external electric fields, and a violation of charge conservation in the presence of certain strain fields. 

To motivate this anomaly, let us first choose a direction $\hat{{\bf{k}}}_0$ in momentum space, and consider a section of a generic 2D Fermi surface  that satisfies $({\bf{v}}_{{\bf k}_{FS}}\cdot\hat{{\bf{k}}}_0) \neq 0,$ where ${\bf{v}}_{{{\bf k}}_{FS}}$ is the Fermi velocity at momentum ${\bf{k}}_{FS}$ on the Fermi surface.  
We can associate an axial current to each quasi-1D  system at constant $k_\parallel \equiv ({\bf k} \times \hat{{\bf{k}}}_0) \cdot \hat{\bf z}$, via a momentum-resolved chirality $\chi^{\hat{\bf k}_0}_{k_{\parallel}}=\sum_{{\bf k}_{FS}}{\text{sgn}}({\bf{v}}_{{\bf k}_{FS}}\cdot\hat{{\bf{k}}}_0)$, where the sum runs over momenta on the relevant section of the Fermi surface for which $({\bf k}_{FS} \times \hat{{\bf{k}}}_0) \cdot \hat{\bf z}=k_\parallel $. For example, if we fix ${\hat{\bf{k}}}_0=\hat{y}$, then at fixed  $k_x=K_x$, we can define an axial current 
$
    j^0_{\chi,y} (K_x)  = \sum_{k_{y, FS}} \chi^{\hat{y}}_{K_x} \hat{n}_{(K_x,k_{y,FS})}; 
$ choosing ${\hat{\bf{k}}}_0=\hat{x}$ we can define $j^0_{\chi,x}(K_{y})$ similarly. Intuitively, this definition captures the fact that if we pick a direction $\hat{{\bf{k}}}_0$ that intersects the Fermi surface going from occupied states to unoccupied states then it has a chiral dispersion along $\hat{{\bf{k}}}_0,$ and vice-versa for anti-chiral.

From a global perspective, any closed Fermi surface will intersect each slice at fixed $k_{\parallel}$ an even number of times along the $\hat{\bf{k}}_0$ direction, with equal numbers of positive and negative chirality intersections. 
Thus for closed Fermi surfaces the total chirality of each fixed $k_\parallel$ slice vanishes. In contrast, Eq. \ref{kxky} describes open Fermi surfaces which have well-defined, non-vanishing chiralities 
$\chi^{\hat{x}}_{K_y}=\text{sgn}(v\xi K_{y})$ 
for slices at fixed $k_y=K_y$, and similarly for fixed $k_x=K_x.$  These values are non-vanishing on each hyperbolic branch of the Fermi surface, since there is only one value of $k_i$ on the Fermi surface at which there is an intersection with each constant $\sigma^{ij}k_{j}$ slice (see Fig. \ref{fig:3dplots_rnak-12}(b),(c)). Thus, each fixed momentum slice is chiral. Despite this chirality, Eq. \ref{kxky} has time-reversal symmetry, which implies $\chi^{\hat{{\bf{k}}}_0}_{-k_{\parallel}} = -\chi^{\hat{{\bf{k}}}_0}_{k_{\parallel}} $, and hence the net axial charge of the entire Fermi surface vanishes, since each branch of the hyperbola has an opposite chirality. This confirms our claim above about the lack of a conventional axial anomaly in this system. 

Interestingly, this failure points the way to the actual anomaly of interest since the \emph{product} $\chi^{\hat{{\bf{k}}}_0}_{k_{\parallel}}\cdot k_{\parallel}$ does take the same sign on the two hyperbolic Fermi surface branches, and will also do so in general for a Fermi surface interval and its time-reversed partner. Hence, let us specialize to time-reversal invariant systems, and focus on anomalies in the momentum densities/currents
\begin{equation} \label{MomentumCharges}
    \hat{\mathcal{J}}^{0}_a=  \frac{1}{A_{2D}}\sum_{\textbf{k}}\hbar k_a\hat{n}_\textbf{k}, 
\end{equation} where $A_{2D}$ is the area of the 2D system.  More precisely, given an interval of the Fermi surface with a non-vanishing chirality $\chi^{\hat{{\bf{k}}}_0}_{k_{\parallel}},$ and its time-reversed partner, we can apply a uniform electric field in the $\hat{{\bf{k}}}_0$-direction, and consider the change in the $k_{\parallel}$ momentum, i.e., the component of $\hat{\mathcal{J}}_a^0$ perpendicular to the applied field. Physically we expect that an electric field acting on a Fermi surface with a non-vanishing $\chi^{\hat{{\bf{k}}}_0}_{k_{\parallel}}$ generates electrons with one sign of $k_{\parallel}$ momentum, and its time-reversed partner will remove electrons with the opposite sign of $k_{\parallel}$ momentum, such that the particle number stays fixed, but there is a net change in momentum. For the hyperbolic Fermi surfaces of Eq. \ref{kxky} we find that, for any value of $\mu$, turning on $E_x$ by adiabatically shifting $A_x$ by $\tfrac{h}{eL}$ generates a change in the $y$-momentum density equal to $\Delta \mathcal{J}_{y}^{0}=-{\text{sgn}}(v\xi)\hbar\sum_{k_y=-\Lambda_y}^{\Lambda_y} |k_y|,$ where $\Lambda_y$ is a wavevector cutoff. If we repeat the experiment with $E_y$ we find $\Delta \mathcal{J}_x^0=-{\text{sgn}}(v\xi)\hbar\sum_{k_x=-\Lambda_x}^{\Lambda_x}|k_x|.$ In the thermodynamic limit, we find $\Delta \mathcal{J}_0^a=-{\text{sgn}}(v\xi)\tfrac{2\Lambda_a }{2\pi}\tfrac{\hbar  \Lambda_a}{2}$ when an electric field $\sigma^{ab}E_b$ is applied by inserting one flux quantum. There are similar anomalous responses of the Eq. \ref{kxky} Fermi surface for momenta orthogonal to other applied electric field directions, except when the field is applied along the $\hat{x}\pm\hat{y}$ directions where the anomalous response vanishes.

Naively we would like to associate the sign of the momentum anomaly response $\text{sgn}(v\xi)$ to a notion of two-dimensional chirality. However, without additional symmetry, this chirality is not well-defined since by simply rotating the coordinate system one can transform $k_x k_y\to -k_x k_y$, hence flipping the sign of the chirality.  Thus, to formulate a robust notion of a rank-2 chirality we need to impose symmetry. Let us impose mirror symmetry about the line $x=y.$ This accomplishes several things: (i) it forces $k_xk_y$ and $k_x^2-k_y^2$ to lie in different symmetry representations such that one can no longer continuously deform $k_x k_y\to-k_x k_y$ without breaking symmetry\footnote{The mirror symmetry does allow for the addition of $\beta (k_x^2+k_y^2),$ to the dispersion, but as long as this term is not large enough to drive a Lifshitz transition that disrupts the hyperbola-like Fermi surface (i.e., let $|\beta|<\tfrac{1}{2}|\hbar v\xi|$), then the chirality remains well defined.}, (ii) it establishes a natural pair of directions (up to a sign) in which to consider the anomalous momentum response; these directions are exchanged by the mirror symmetry, and orthogonal to each other, e.g., $\hat{x},\hat{y}$ for this mirror symmetry, (iii) it requires $\Lambda_x = \Lambda_y \equiv \Lambda.$  
Hence, with mirror symmetry the anomalous momentum response is characterized by:
\begin{equation}
    \partial_\mu  \mathcal{J}_{a}^{\mu}=\chi_{2} \sigma_{ab} \frac{e \Lambda^2}{4\pi^2} E_b\,\,\, a,b =x ,y,\label{eq:rank2panomaly}
\end{equation}
where $\chi_2$ is the rank-2 chirality, which is equal to $\text{sgn}(v\xi)$ for the dispersion in Eq. \ref{kxky}.

With mirror symmetry we now have a well-defined notion of a chirality, but we still have not indicated a link to a rank-2 structure. The rank-2 nature is more clearly expressed at this level through a reciprocal anomalous response where a gauge field $\mathfrak{e}_\mu^a,$ that couples canonically to the momentum current $\mathcal{J}_a^\mu,$ can produce an anomaly in the ordinary charge current. 
Eq. (\ref{MomentumCharges}) implies that each electron couples to  $\mathfrak{e}_\mu^a$ with a charge equal to its $a$-momentum; thus adiabatically shifting $\mathfrak{e}^a_j$ is heuristically like turning on an electric field in the $-j$ ($+j$)-direction for electrons with negative (positive) $k_a$ momentum, with a magnitude proportional to $|k_a|$.  
For the rank-2 chiral fermions, the opposite electric fields are applied to opposite chirality Fermi surface branches, hence we expect such a shift will generate an excess of charge for the rank-2 chiral fermions. We note in passing that while it is tempting to identify the fields $\mathfrak{e}_\mu^a$ as frame fields, we only consider fields that couple to momenta along translationally invariant lattice directions, which has the effect of limiting the gauge transformations to $\delta_{\boldsymbol{\lambda}}\mathfrak{e}_\mu^a=\partial_\mu \lambda^a.$  

To see the anomaly explicitly, let us shift $\mathfrak{e}_x^y$ by $L_y/L_x$ in a system with periodic boundary conditions.  Physically, this describes a process in which a dislocation with Burgers vector $L_y\hat{y}$ is threaded adiabatically through the hole spanned by the periodic $x$-direction\cite{hughesleighfradkin} (the final result is akin to a twisted carbon nanotube).  
Mathematically, for an electron having a fixed $k_y$, this shift is equivalent to a shift in $A_x$ by $\tfrac{\hbar}{e}k_y\mathfrak{e}_x^y$\cite{hughesleighfradkin,parrikar1,parrikar2}.  Hence an electron with momentum $k_y=2\pi s_y/L_y$ ($s_y\in\mathbb{Z}$) will experience a shift $k_x\to k_x +2\pi s_y/L_x.$  If the states at this $k_y$ have $\chi^{\hat{x}}_{k_{y,FS}}>0$ ($\chi^{\hat{x}}_{k_{y,FS}}<0$)  then this will generate $s_y$ ($-s_y$) particles at $k_y.$ The states at $-k_y$ have the opposite chirality, but also the opposite shift, and thus also contribute $s_y$ ($-s_y$) particles. In total the change in the charge for positive chirality is $\Delta Q=e\sum_{k_y=-\Lambda_y}^{\Lambda_y}\tfrac{L_y}{2\pi}k_y=e\left(\frac{L_y \Lambda_y}{2\pi}\right)^2.$  In the presence of mirror symmetry there is a symmetric effect for a shift in  $\mathfrak{e}_y^x\to\mathfrak{e}_y^x+L_x/L_y$ that will produce a $\Delta Q=e \left(\frac{L_x \Lambda_x}{2\pi}\right)^2,$ and these effects can be summarized by an anomalous conservation law 
\begin{equation} \label{eq:rank2anomalyaction}
 \partial_\mu j^\mu=\chi_2 \frac{e\Lambda^2 }{4\pi^2}\mathcal{E}_{xy},
\end{equation} where mirror fixed $L_x=L_y=L, \Lambda_x=\Lambda_y=\Lambda,$ and $\mathcal{E}_{xy}=\partial_x \mathfrak{e}_0^y+\partial_y\mathfrak{e}_0^x-\partial_t\mathfrak{e}_{xy}$ is an effective rank-2 electric field (with a vector charge Gauss' law\cite{rasmussen16,pretko17,pretko18}). Here, the mirror symmetry allows us to combine $\mathfrak{e}_{xy}\equiv \mathfrak{e}_{x}^y+\mathfrak{e}_{y}^x$ to form a symmetric rank-2 gauge field with gauge transformations $\mathfrak{e}_{xy}\to\mathfrak{e}_{xy}+\partial_x \lambda_y+\partial_y\lambda_x$ which match those of a vector-charge rank-2 gauge field where the vector charge is crystal momentum. Hence, our mirror-symmetric rank-2 chiral fermions have an anomalous rank-2 response, which is the origin of their name.

Before moving on to lattice models with rank-2 chiral fermions, a few remarks in order. First, for more generic dispersion relations $E=g^{ij}k_i k_j,$ where $g^{ij}$ is a \emph{Lorentzian} metric, a rank-2 chirality can be defined by choosing a mirror line $M$ through the origin of momentum space, i.e., $M_{\gamma_1,\gamma_2}:\gamma_x k_x=\gamma_y k_y,$ that exchanges the principal axes $\hat{x}_1,\hat{x}_2$ of the quadratic form $g^{ij}.$ These axes are unique up to a sign, which 
 will preserve the relative distinction between the two chiralities. 
The resulting rank-2 chiral fermions 
exhibit anomalies described by Eqs. \ref{eq:rank2panomaly} and \ref{eq:rank2anomalyaction}, but with the replacement  $\mathcal{E}_{xy}\to\mathcal{E}_{x_1 x_2}$ etc. A second comment concerns the cutoff dependence of the response action.  We emphasize that this is a momentum cutoff, {\it not} an energy cutoff: it describes the range of ${\bf k}$ values over which we have included states on the Fermi surface.\footnote{While for a continuum theory this might seem problematic it is not completely unexpected\cite{hughesleighfradkin,parrikar1,bradlynrao} as momentum transport can be dominated by, and diverge due to, the states at large momentum, which hence have to be cut off. }  Below, we consider a 3D  model where these cutoffs have a natural physical origin.

In 2D, where the Fermi surface must be closed, the net anomalous response necessarily vanishes when we extend the momentum cutoff to include the entire Brillouin zone.  
To illustrate this, consider a 1-band, 2D lattice model on a square lattice with only next-nearest neighbor hopping such that the dispersion relation is given by $E({\bf{k}})=\sin k_x \sin k_y.$  This model has $M_{1,1}$ mirror symmetry, hence we  evaluate the $k_{x}$ and $k_{y}$ momentum anomalies.  If one expands around $(k_x,k_y)=(0,0)$ the dispersion is $E({\bf{k}})\sim \chi_2 k_x k_y$ with $\chi_2 = +1$, while around $(k_x,k_y)=(\pi,0)$ the dispersion is $E({\bf{k}})\sim \chi_2' k_x k_y$ , with $\chi_2'=-1$ (see Fig. \ref{fig:3dplots_rnak-12}(d)).   Hence, we find an equal number of chiral and anti-chiral modes transverse to $k_x$ (and to $k_y$), which make opposite contributions in Eq. (\ref{eq:rank2panomaly}).

\begin{figure}
    \centering
    \includegraphics[width=0.45\textwidth]{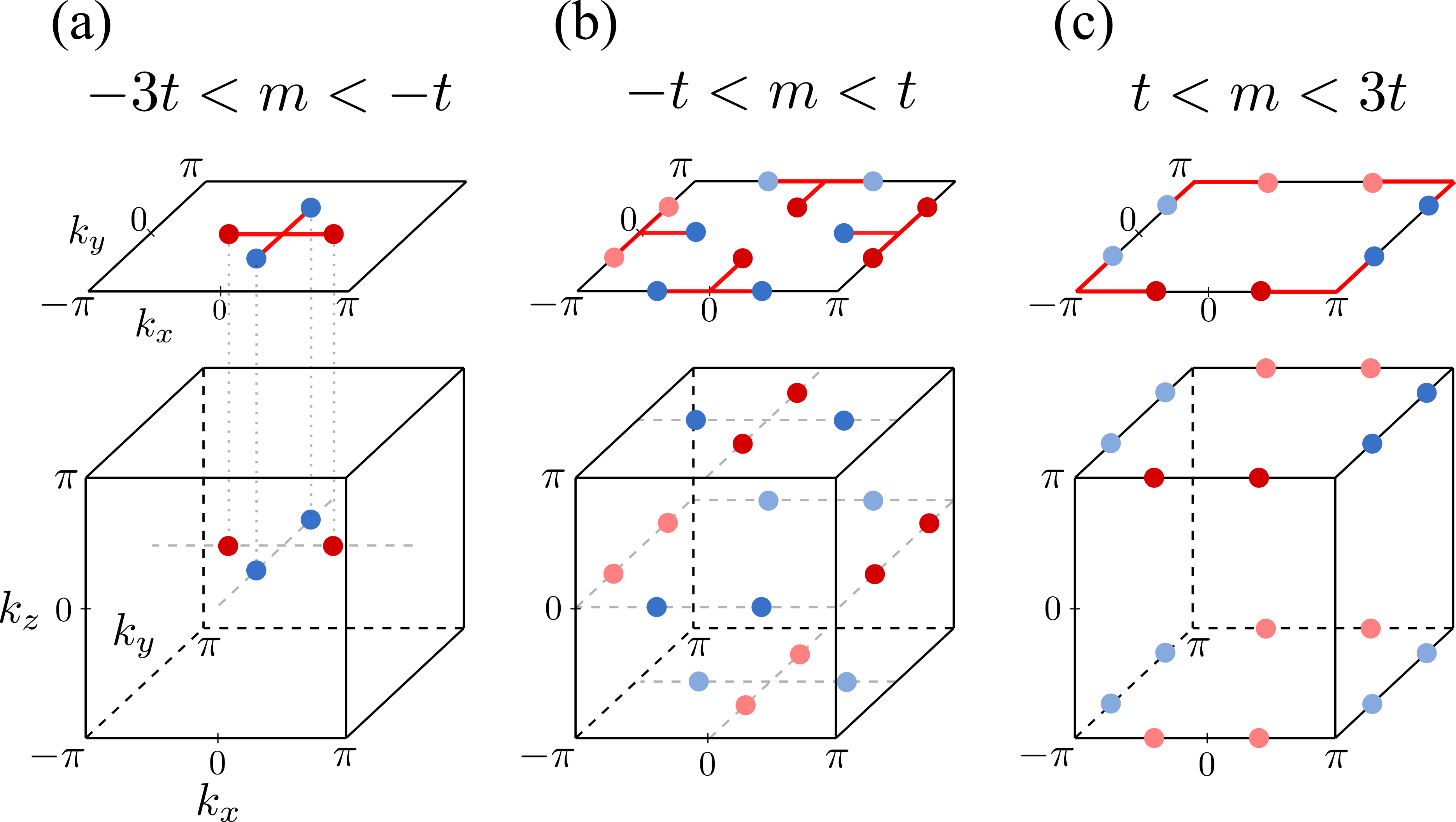}\\
    \includegraphics[width=0.45\textwidth]{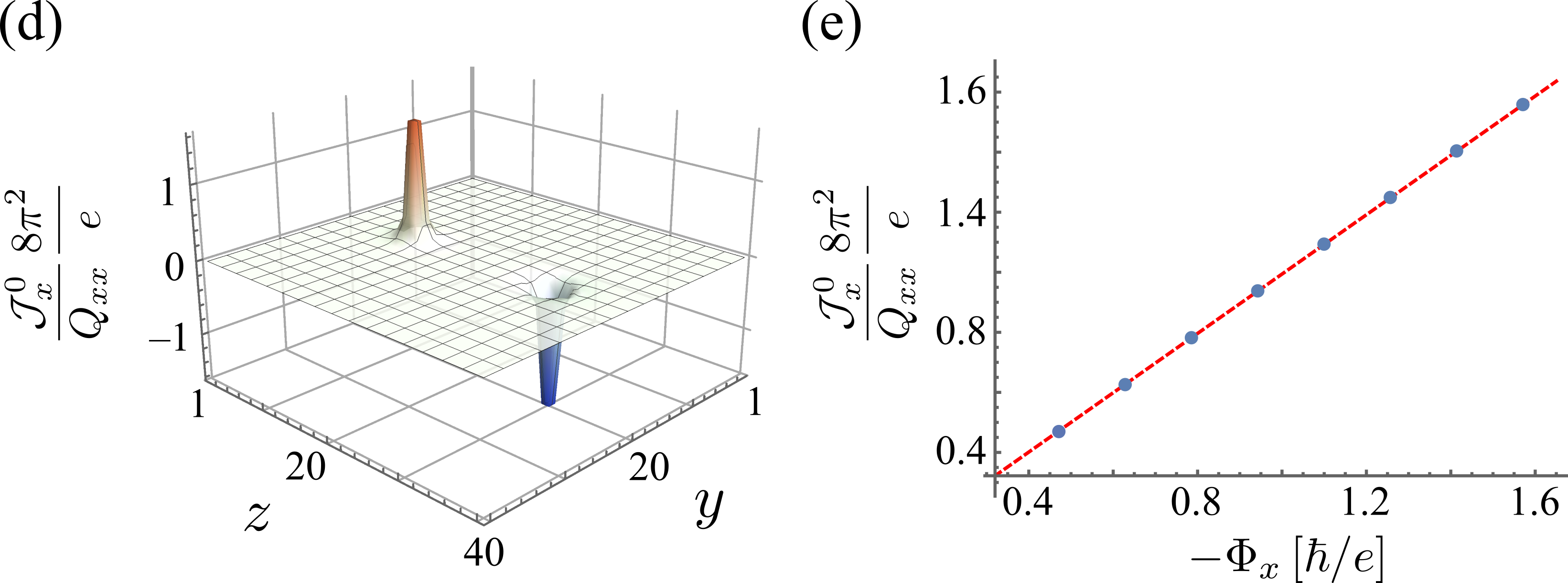}
    \caption{(a)--(c): Weyl node positions in the 3D bulk Brillouin zone and surface state structure on the (0,0,1) surface for different values of the parameters $t$ and $m$. Red (blue) dots depict Weyl nodes with  chirality $\chi=+1 (-1).$  Red lines in the 2D surface BZ depict the surface zero modes. (d): Distribution of the $k_x$ momentum density $\mathcal{J}^0_x$ weighted by the Weyl quadrupole moment $Q_{xx}=\pi^2/2$ in units of $e/8\pi^2$ for a lattice with $N_y\times N_z=40\times 40$ and $m=-2t$ in the presence of two opposite magnetic flux lines carrying $\Phi_x=\pm 1/3$ magnetic flux quanta at $(y,z)=(20,10)$ and $(y,z)=(20,30)$ respectively (see Supplement for details). (e): Dependence of the momentum density localized in the vicinity of magnetic flux line at $(y,z)=(20,10)$ on the amount of magnetic flux $\Phi_x$ threaded through this plaquette. Red line is a fit of the numerical data (blue dots) with the slope $\approx 0.98795$ which demonstrates that the prefactor of the response action is in agreement with the expected value of $e/(8\pi^2)$.
    We find the same dependence for the charge density $j^0$ localized on a torsional magnetic flux $\mathcal{B}^x_x$ (see Supplement for Figure).}
    \label{fig:dxky/dt}
\end{figure}

A net rank-2 chirality can be realized, however, at the surface of a 3D system.
Let us consider a 2-band Bloch Hamiltonian for a Weyl semimetal:
\begin{equation}
\begin{split}
    H(\textbf{k})=&\sin k_x \sin k_y \Gamma^x+\sin k_z \Gamma^y+\\
    &+\left(m+t(\cos k_x+\cos k_y+\cos k_z) \right)\Gamma^z,
    \label{eqn:ham_kxky}
\end{split}
\end{equation}
where if $\Gamma^a=\tau^a$ Pauli matrices the model has time-reversal symmetry with $T=K$, $C_2^z=\mathbb{I},$ and $M_{1,1}=\mathbb{I},$ or if $\Gamma^x=\tau^y\otimes\sigma^y,\Gamma^y=\tau^x\otimes\sigma^y,\Gamma^z=\tau^z\otimes\mathbb{I}$ the model has spinful time-reversal symmetry $T=i\mathbb{I}\otimes\sigma^y K.$ We focus on the two-band model for simplicity, as the four-band model behaves just as two copies of the former. In this model we find several Weyl semimetal regimes summarized in Fig. \ref{fig:dxky/dt}(a),(b),(c). In particular, let us focus on the range  $-3t<m<-t$, where the system has four gapless Weyl points in the $k_z=0$ plane located at: $\textbf{k}=(\pm\arccos(-m/t-2),0,0)^T$ and $(0,\pm\arccos(-m/t-2),0)^T$. 
One can establish the existence of rank-2 chiral fermions on the $z$-surfaces via numerical diagonalization, or with analytical lattice methods to solve for surface states shown in, for example, Refs. \onlinecite{konig2008,jpsj,vatsal}. As indicated by the surface BZ projections in Fig. \ref{fig:dxky/dt}, we find  surface states that have a dispersion relation $E({\bf{k}})=\pm\sin k_x \sin k_y$ centered around the $\Gamma$-point, with zero-energy lines that terminate at the four Weyl nodes, and an overall sign that flips when the surface normal vector is $\pm \hat{z}$. On the side surfaces we find Fermi arcs that  exhibit an anisotropic momentum anomaly, but which are not rank-2 chiral fermions.

From our continuum calculations we expect the rank-2 surface states to be anomalous, with the momentum locations of the Weyl nodes serving as natural momentum cutoffs in the $x$ and $y$-directions. The remaining question is: can we describe the anomalous surface response as a bulk response in analogy to how the anomalous response of chiral Fermi arcs encode the bulk anomalous Hall effect\cite{burkov2011A,Wan2011,zyuzin2012,RamamurthyPatterns}. To this end, let us consider the momentum space locations of the Weyl nodes ${\bf{K}}^{(\alpha)}$. We find that the pair of nodes lying on the $k_x$ axis have Weyl chirality $+1$, while the pair of nodes on the $k_y$ axis both have Weyl chirality $-1.$ Hence, the total chirality and Weyl momentum dipole moment (i.e., the anomalous Hall coefficients) vanish: $\sum_{\alpha=1}^{4}\chi_{(\alpha)}=0,$ $\sum_{\alpha=1}^{4}{\bf{K}}^{(\alpha)}\chi_{(\alpha)}=0.$ However, we note that the Weyl momentum quadrupole moment is non-vanishing:
\begin{equation}
    Q_{xx}=\sum_{\alpha=1}^{4}\chi_{(\alpha)}(K^{(\alpha)}_x)^2=-\sum_{\alpha=1}^{4}\chi_{(\alpha)}(K^{(\alpha)}_y)^2=Q_{yy}.
\end{equation} Indeed since the mirror symmetry $M_{(1,1)}$ also flips the sign of the Weyl chiralties it enforces $Q_{xx}=-Q_{yy},$ and $Q_{xy}=0.$ We will see below that the anomalous rank-2 surface response is determined by exactly this Weyl quadrupole moment.

To determine the bulk response we calculate a charge-current ($\hat{j}^i=\partial_{k_i}E(k)$)-- momentum-current ($\hat{\mathcal{J}}^i_a=k_a \hat{j}^i)$ correlation function using the Kubo formula (see Supplement). We find a bulk linear response:
 \begin{eqnarray}
    \mathcal{J}_{a}^{\mu}&=&\frac{e}{8\pi^2}\epsilon^{\mu\nu\rho\sigma}Q_{\nu a}\partial_\rho A_\sigma ,\nonumber\\
    j^{\mu}&=&\frac{e}{8\pi^2}\epsilon^{\mu\nu\rho\sigma}Q_{\nu a}\partial_\rho \mathfrak{e}_{\sigma}^a,\label{eq:responsecurrents}
    \end{eqnarray} where (when $\mu=0$)  the response coefficients are:
\begin{equation}
    Q_{ia}=-\frac{1}{2\pi}\int d^3\textbf{k}\ \epsilon_{ijk} k_a \mathcal{F}^{jk},
    \label{eqn:resp_coeff}
\end{equation}
where $\mathcal{F}^{ij}$ 
is the Berry curvature. We show in the Supplement that for a gapless system, $Q_{ia}$ can be reduced to an integral over the Fermi surface analogous to the arguments in Ref. \onlinecite{haldaneAHE}:
\begin{equation}
    Q_{ia}=\frac{1}{2\pi}\int_{FS} k_a k_i \mathcal{F}^{\mu\nu} \ ds_\mu \wedge ds_\nu,
    \end{equation}
where $\{s_1,s_2\}$ are the coordinates that parametrize the Fermi surface. 
For a Weyl semimetal, the Fermi surface splits into disjoint Fermi surfaces $FS_\alpha,$ each enclosing an individual Weyl node carrying a Weyl chirality $\chi_{(\alpha)}.$ Hence, the integral over the Fermi surface simplifies to a sum over a discrete set of Weyl nodes:
\begin{equation}
    Q_{ia}=\sum_{\alpha=1}^{N_{\text{Nodes}}} \chi_{(\alpha)} K_a^{(\alpha)} K^{(\alpha)}_i,
\end{equation} and is exactly the Weyl momentum space quadrupole moment mentioned above. We note that recent work has shown that the Hall viscosity can be determined by the momentum space quadrupole of the occupied states\cite{bradlynrao}, while our work shows that a similar quantity,  calculated on only the Fermi surface, describes a mixed charge-momentum response.

The bulk responses in Eq. (\ref{eq:responsecurrents}) are described by the action
\begin{equation}
S[A,\mathfrak{e}]=-\frac{e}{8\pi^2}\int d^4 x  \epsilon^{i\nu\rho\sigma}Q_{i a}A_\nu \partial_\rho \mathfrak{e}_\sigma^a . \label{eq:genbulkresponse}
\end{equation}
 For our model, and any Weyl quadrupole model having time-reversal and $M_{(1,1)}$  mirror symmetry,  we find a single non-vanishing response coefficient: $\bar{Q}=Q_{xx}=-Q_{yy}$  (note $Q_{xy}=0$ by symmetry), and we can simplify the response theory to find:
\begin{equation}
    S=-\frac{e\bar{Q}}{8\pi^2}\int d^4 x\left[A_z \mathcal{E}_{xy}-A_y \mathcal{E}_{z}^x-A_x \mathcal{E}_{z}^y-A_0(\mathcal{B}_y^y-\mathcal{B}_x^x)\right],\label{eq:momentumchernsimons}
\end{equation} where $\mathcal{E}_z^a=\partial_z \mathfrak{e}_0^a-\partial_t \mathfrak{e}_z^z, $ and $\mathcal{B}_i^a=\epsilon^{ijk}\partial_j \mathfrak{e}_k^a.$ We write the action in this suggestive form to make a direct comparison to the rank-2 scalar charge dipole Chern-Simons theory of Ref. \onlinecite{ybh2019}. Indeed we see Eq. \ref{eq:momentumchernsimons} is just a vector-charge version of that Chern-Simons theory, i.e, instead of a rank-2 scalar dipole charge we have a rank-2 vector momentum charge. Interestingly, unlike the scalar charge case which was shown to be a pure boundary term\cite{ybh2019}, this theory gives the bulk responses in Eq. \ref{eq:responsecurrents}. This action should be contrasted with previous work on geometric response in Weyl semimetals which focus on the terms induced by a Weyl dipole\cite{you2016,Huang19,Ferreiros19,Liang20}. 

Let us now try to understand the physical meaning of the bulk response. First, consider
    \begin{equation}
        \mathcal{J}_{a}^0=\frac{e\bar{Q}}{8\pi^2}B_a (\delta_{ax}-\delta_{ay}),
    \end{equation} which indicates a momentum density attached to magnetic flux. Second, consider 
  \begin{equation}
        j^0=\frac{e\bar{Q}}{8\pi^2}\mathcal{B}_{i}^a(\delta^{ix}\delta_{ax}-\delta^{iy}\delta_{ay}),
    \end{equation}\noindent which is a charge density attached to a torsional magnetic field $\mathcal{B}^a_{i}$ parallel to the $i$-th direction and having Burgers vector in the $a$-th direction. For our model, the second response requires the torsional flux and Burgers vector to be parallel, which is naturally represented by a screw dislocation. Hence, on screw dislocations in the $x$ or $y$-directions we should find a bound charge density. We numerically confirm both of these responses by inserting localized magnetic (torsional) flux tubes in the $x$-direction and calculating the localized momentum (charge) density. Our numerical results exactly match the response equations as shown in Fig. \ref{fig:dxky/dt}(c),(d).
   
To see how this bulk response is connected to the surface states, we observe that the response action Eq.  \ref{eq:momentumchernsimons} is not gauge invariant in the presence of a boundary. As a model for the surface we can let $\bar{Q}(z)$ be a domain wall in the $z$-direction\cite{RamamurthyPatterns}. If we put such a domain wall configuration into Eq. \ref{eq:momentumchernsimons} and integrate by parts, then the action localized on the domain wall is
    \begin{equation}
   S_{\partial z}=     \frac{e\bar{Q}}{8\pi^2}\int dx dy dt\left[A_0\mathfrak{e}_{xy}-A_y\mathfrak{e}_0^x-A_{x}\mathfrak{e}_0^y\right].
    \end{equation} If we treat the Weyl points as momentum cutoffs in the $x$ and $y$ directions then we can make the replacement $\bar{Q}=2\Lambda^2.$ The end result has response equations that exactly match the anomalous conservation laws in Eqs. (\ref{eq:rank2panomaly}) and (\ref{eq:rank2anomalyaction}). Thus the rank-2 chiral fermions are the boundary manifestation for the bulk vector-charge Chern-Simons response action.

This work has shown that rank-2 chiral fermions, associated with 2D time-reversal and mirror-symmetric systems, exhibit a new type of anomaly, in which momentum transverse to the direction of an applied electric field is not conserved, and where certain lattice defects also lead to violations of charge conservation. We expect that rank-2 chiral fermions can naturally appear in time-reversal invariant Weyl semimetals having non-vanishing Weyl quadrupoles.  These results represent the first in a family of higher-dimensional systems with various notions of higher-rank chirality. We elaborate on other members of this family in a forthcoming work.

\acknowledgements O.D. and T.L.H. thank  ARO MURI
W911NF2020166 for support. F. J. B. is grateful for the financial support of NSF DMR 1928166, the Carnegie Corporation of New York, and the Institute for Advanced Study.

\bibliography{References}

\appendix

\section{Kubo Formula}
In this section we present a detailed derivation of the response equations (\ref{eq:responsecurrents}). 
For clarity, we will limit the derivation here to a 2-band free-fermion model with the Hamiltonian of the general form $H(\textbf{k})=\textbf{d}(\textbf{k})\cdot\boldsymbol{\sigma}.$ We will eventually specialize to a case with a band structure in which Weyl nodes are arranged in a quadrupolar pattern. 
Let us consider the mixed momentum current $\mathcal{J}^\mu_a$ -- electromagnetic current $j^\lambda$ response and calculate the corresponding linear response coefficient:
\begin{equation}
    \mathfrak{t}^{\mu,\lambda}_a=\lim_{\omega\to 0}\frac{i}{\omega}\Pi^{\mu,\lambda}_a(\omega+i\delta),
\end{equation}
\begin{equation}
\begin{split}
    \Pi^{\mu,\lambda}_a(i\nu_m)=\frac{1}{V\beta}\sum_{\textbf{k},n}\text{tr}\big[&\mathcal{J}^\mu_a(\textbf{k}) G(\textbf{k},i(\omega_n+\nu_m))\\
    &j^\lambda(\textbf{k})G(\textbf{k},i(\omega_n))\big].
\end{split}
\label{eqn:stress-curr-corr}
\end{equation}
The electromagnetic current carried by a single-particle states with momentum $\textbf{k}$ is defined in the usual way:
\begin{equation}
    j^\lambda (\textbf{k})=i\frac{e}{\hbar}\langle [\hat{H},\hat{x}^\lambda]\rangle_{\textbf{k}}=\frac{e}{\hbar}\frac{\partial H(\textbf{k})}{\partial k_\lambda}.
\end{equation}
Assuming that the internal degrees of freedom transform trivially under rotations we define unsymmetrized strain generators as 
$\hat{J}^\mu_a=\hat{p}_a\hat{x}^\mu$ which lead to the following expression for the momentum currents tensor:
\begin{equation}
    \mathcal{J}^\mu_a(\textbf{k})=-\frac{i}{\hbar}\langle[\hat{H},\hat{J}^\mu_a]\rangle_{\textbf{k}} =k_a\frac{\partial H(\textbf{k})}{\partial k_\mu}\equiv \frac{\hbar}{e}k_a j^\mu(\textbf{k}).
\end{equation}
From this we recognize that Eq. \ref{eqn:stress-curr-corr} is a conventional current-current response function, but with an additional factor of $k_a$ in the summand.
Following the same steps as in the derivation of the Hall conductivity for a 2-band free-fermion model in Ref. \onlinecite{Qi2006}, we arrive at the following equation for our linear response coefficient:
\begin{equation}
\begin{split}
        \mathfrak{t}^{\mu,\lambda}_a=\frac{1}{2V}\sum_{\textbf{k}}&k_a \left[\epsilon^{\alpha\beta\gamma}\frac{\partial \hat{d}_\alpha(\textbf{k})}{\partial k_\mu}\frac{\partial \hat{d}_\beta(\textbf{k})}{\partial k_\lambda}\hat{d}_\gamma(\textbf{k})\right]\\
        &\times\left(n_+(\textbf{k})-n_-(\textbf{k})\right).
        \label{eqn:lin_resp}
\end{split}
\end{equation}
where $n_+(\textbf{k})$ $(n_-(\textbf{k}))$ is the occupation of states at momentum ${\textbf{k}}$ above (below) the Fermi level.
Setting the Fermi level precisely at the Weyl nodes, which fills up completely the lower band and leaves the upper band empty, and translating from the sum over momentum space to a continuous integral, we find:
\begin{equation}
    \mathfrak{t}^{\mu,\lambda}_a=-\frac{1}{(2\pi)^3}\int_{FV} d^3\textbf{k}\ k_a \mathcal{F}^{\mu\lambda}
\end{equation} where $\mathcal{F}^{\mu\lambda}$ is the Berry curvature and the integral is taken over the Fermi Volume.
The combination in the integrand has a very intuitive interpretation: for particular choices of response coefficients one can rewrite this integral as an integral over the Fermi surface. For example, we can show that:
\begin{equation}
\begin{split}
    \mathfrak{t}^{y,z}_x &=-\frac{1}{(2\pi)^3}\int_{FV} d^3\textbf{k}\ k_x\mathcal{F}^{yz}\\
    &=-\frac{1}{(2\pi)^3}\int_{FV}d\left(k_xA^z dk_z\wedge dk_x - k_x A^y dk_x\wedge dk_y\right).
\end{split}
\end{equation}
Since the integrand in this expression is a total derivative, we can reduce the expression for $\mathfrak{t}^{y,z}_x$ to the integral over the Fermi Surface. Furthermore, introducing a pair of coordinates $\{s_1,s_2\}$ parametrizing the Fermi Surface we get:
\begin{equation}
\begin{split}
    \mathfrak{t}^{y,z}_x&=-\frac{1}{(2\pi)^3}\int_{FS} k_xA^z dk_z\wedge dk_x - k_x A^y dk_x\wedge dk_y\\
    &=-\frac{1}{2(2\pi)^3}\int_{FS} k^2_x(\textbf{s})\mathcal{F}^{ij}(\textbf{s})ds_i\wedge ds_j
\end{split}
\end{equation}
where $\mathcal{F}^{ij}$ is the Berry curvature expressed in the coordinates on the Fermi Surface, and we have integrated by parts in passing from the first line to the second, giving an extra factor of $k_x$.
Setting the Fermi energy to match the location of the Weyl nodes, the integral over the Fermi surface turns into a sum over the total Berry curvatures localized near each of the $N_{Weyl}$ nodes weighted by the squared $k_x$ position of the node inside the Brillouin zone:
\begin{equation}
    \mathfrak{t}^{y,z}_x=-\frac{1}{8\pi^2}\sum_{\alpha=1}^{N_{Weyl}}\chi_{(\alpha)}(K^{(\alpha)}_x)^2=-\frac{Q_{xx}}{8\pi^2},
\end{equation}
where $\alpha$ is the Weyl node's index and $K^{(\alpha)}_x$ is the position of that node along the $k_x$ axis.
Similarly, we can look at the other coefficients for the $\{\hat{x},\hat{y}\}$ plane momentum current responses to the electric $E_z$ field to find:
\begin{equation}
\begin{split}
    \mathfrak{t}^{x,z}_{y}=\frac{Q_{yy}}{8\pi^2},\quad \mathfrak{t}^{x,z}_{x}-\mathfrak{t}^{y,z}_{y}=-\frac{1}{2(4\pi^2)}Q_{xy},
\end{split}
\end{equation}
where we always define the quadrupole moment components for the set of nodes with chiralities $\chi_{(\alpha)}$ located at the set of $N_{Weyl}$ Weyl points as:
\begin{equation}
    Q_{\mu a}=\sum_{\alpha=1}^{N_{Weyl}}k_\mu k_a \chi_{(\alpha)}.
\end{equation}

Finally, to reconstruct the response action we will use the fact that the coefficient $\mathfrak{t}^{\mu,\lambda}_{a}$ describes the linear response of the $k_a$ momentum current $\mathcal{J}^\mu_a$ to an applied electric field
\begin{equation}
   \mathfrak{t}^{\mu,\lambda}_{a}=\frac{\partial \mathcal{J}^\mu_a}{\partial E_\lambda}=\frac{\partial}{\partial E_\lambda}\frac{\delta S}{\delta \mathfrak{e}^a_\mu}.
\end{equation}
More specifically, from the $k_x$ momentum response to the $E_z$ electric field we find:
\begin{equation}
\begin{split}
    -\frac{Q_{xx}}{8\pi^2}=\mathfrak{t}^{y,z}_{x}&=\frac{\partial}{\partial(\partial_t A_z)}\frac{\delta}{\delta \mathfrak{e}^x_y} \alpha_{ia}\int d^4x \epsilon^{i\nu\rho\sigma}A_\nu\partial_\rho \mathfrak{e}^a_\sigma=\alpha_{xx}.
\end{split}
\end{equation}
Repeating this analysis for other response coefficients allows us to confirm that the matrix $\alpha_{ia}$ in front of the linear response action is given by:
\begin{equation}
    \alpha_{i a}=-\frac{Q_{ia}}{8\pi^2}.
\end{equation}

\section{Numerics}
\label{app:num}
\begin{figure*}
    \centering
    \includegraphics[width=0.95\textwidth]{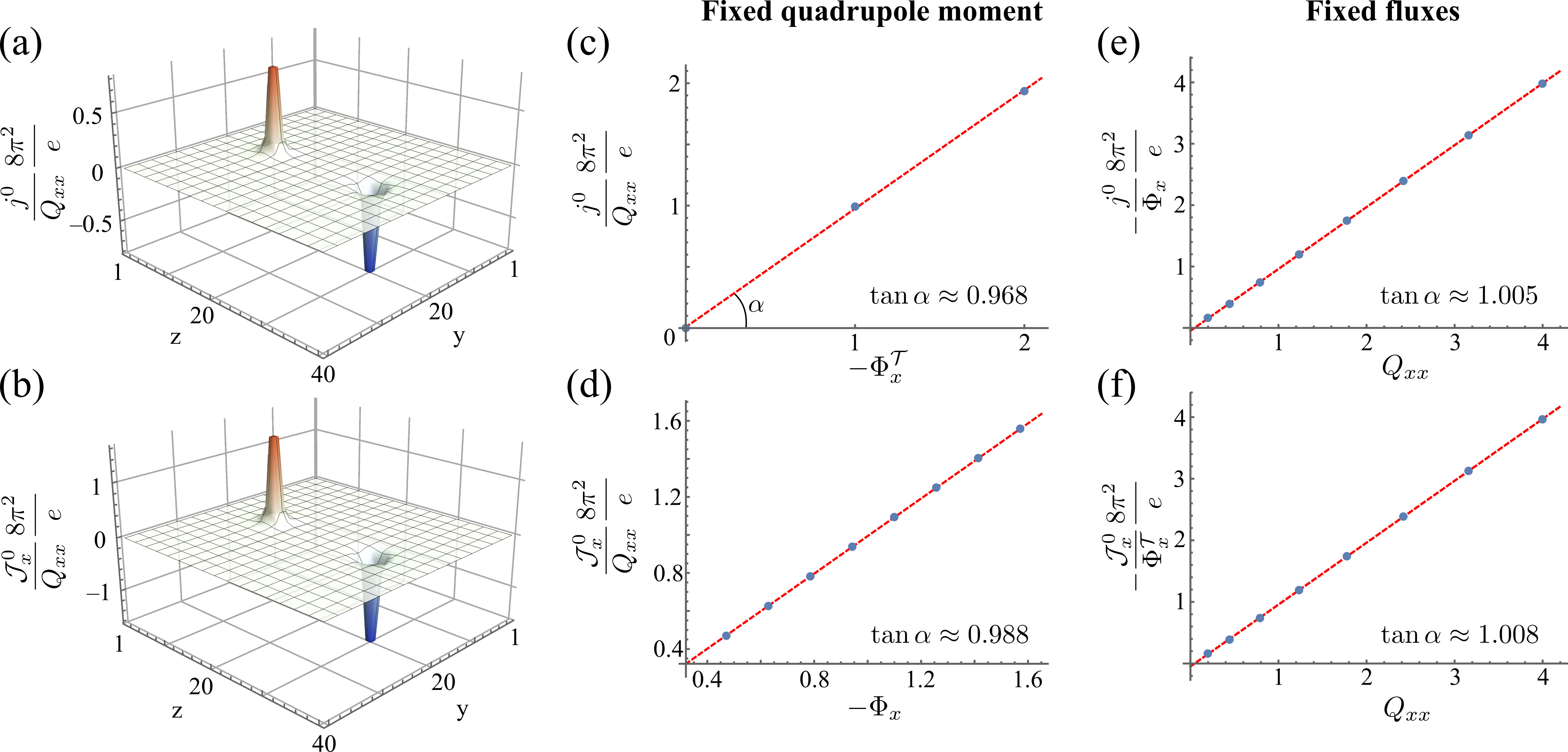}
    \caption{(a): Distribution of charge density in units of the quadrupole moment $Q_{xx}=\pi^2/2$ in the presence of two opposite torsional magnetic flux lines carrying $\Phi^\mathcal{T}_x=\pm 1$ at $(y,z)=(20,10)$ and $(y,z)=(20,30)$ respectively. (b): Distribution of the $k_x$ momentum density in units of $Q_{xx}$ in the presence of two opposite magnetic flux lines carrying $\Phi_x=\pm 1/3$ magnetic flux quanta. In plots (c)-(f) we present the results of each of our numerical simulations (blue dots) along with the linear fit (red dashed lines) for various combinations of parameters.}
    \label{fig:data}
\end{figure*}

In this section we present the results of our numerical simulations which allow us to confirm the structure of the effective response action (\ref{eq:genbulkresponse}). To do this we  will insert a pair of spatially separated lines carrying opposite fluxes (either magnetic or torsional). We study a tight-binding model for a 3D Weyl semimetal with four Weyl nodes arranged in a quadrupolar pattern. 
The particular model we use is given in Eq. \ref{eqn:ham_kxky}. We simulate the model on a completely periodic $N_x\times N_y\times N_z = 40\times 40\times 40$ lattice. We introduce a pair of opposite flux lines (either magnetic or torsional) stretching along the $\hat{x}$-direction, inserted through the plaquettes with coordinates $(y,z)=(20, 10)$ and $(y,z)=(20,30)$. 
We then diagonalize the Hamiltonian, treating the $y$ and $z$-directions in real space and the translationally invariant $x$-direction in momentum space. We  investigate the electric charge and $k_x$-momentum  densities as a function of the $y$ and $z$ coordinates. 
As shown in Fig. \ref{fig:data}, we find a momentum (electric) charge density localized in the vicinity of the (torsional) magnetic flux line. 

We introduce the pair of magnetic flux lines by shifting the Peierls' phases $\alpha_{y}(l)=\frac{e}{\hbar}\int_l A_y dy$  by a constant amount for a set of links $l$ with coordinates $y=20$ and $10\le z< 30$. 
To implement the torsional magnetic flux, we introduce a lattice gauge field $\mathfrak{e}^{a}_{i}$ which couples to the momentum $k_a,$ and then we shift $\mathfrak{e}^x_y$ by a constant amount on the set of links $l$ described above. 
This results in a momentum-dependent Peierls' phase $\beta_y(l)=k_x\int_l \mathfrak{e}^x_y dy$. 
The torsional magnetic flux obtained from shifting $\mathfrak{e}^x_y$ by one lattice constant $a$ is equivalent to introducing a screw dislocation to our system. An electron with fixed momentum $k_x$ in the $\hat{x}$ direction that travels around a torsional flux line  in the $\{\hat{y},\hat{z}\}$ plane results in an effective Aharonov-Bohm phase factor $\text{e}^{ik_x a}$ arising  from translating the electron by one lattice constant in the $\hat{x}$ direction.

To investigate the structure of the effective response action we perform the set four tests: 
(1) we fix the value of the Weyl quadrupole moment to be $Q_{xx}=-Q_{yy}=\pi^2/2$ by setting the model parameters $t=1$, $m=-2t=-2$ and then we track the charge density bound to a torsional magnetic flux line as a function of torsional magnetic flux $\Phi^{\mathcal{T}}x$ (Fig. \ref{fig:data}(c)); 
(2) for the same configuration of the Weyl nodes, we study the dependence of the $k_x$-momentum density bound to a magnetic flux line as a function of magnetic flux $\Phi_x$ (Fig. \ref{fig:data}(d));
(3) we tune the value of the quadrupole moment and track the dependence of the charge density bound to a fixed torsional magnetic field with $\Phi^\mathcal{T}_x=-1$ as a function of the Weyl quadrupole moment's component $Q_{xx}$ which can be varied by tuning the parameters $m$ and $t$ of the tight-binding model (Fig. \ref{fig:data}(e));
(4) similarly, we track the $k_x$-momentum density bound to a magnetic flux line with $\Phi_x=1/N_x$ magnetic flux quanta as a function of the quadrupole moment $Q_{xx}$ (Fig. \ref{fig:data}(f)).
In all four cases we observe linear dependencies with the slopes of all four plots closely approximating the coefficient of $1/(8\pi^2)$ in front of the action.

\end{document}